\documentclass{moriond}

\usepackage{amsmath,amssymb}

\def\be{\begin{equation}}
\def\ee{\end{equation}}
\def\bea{\begin{eqnarray}}
\def\eea{\end{eqnarray}}

\newcommand{\Photo}{}

\begin{document}
\vspace*{4cm}
\title{Flavour hierarchies, extended groups and composites}

\author{Javier M. Lizana}

\address{Instituto de F\'isica Te\'orica UAM/CSIC, Nicolas Cabrera 13-15, Madrid 28049, Spain}

\maketitle\abstracts{
In these proceedings, I present a composite Higgs model in which the flavour hierarchies between the third and light families emerge naturally. 
In particular, CKM mixing angles turn out to be suppressed while PMNS matrix remains anarchic.
This flavour structure arises as a consequence of the extended non-universal gauge symmetry of the model and the electroweak charges of the fundamental fermions of the new composite sector that realises the Higgs boson as a pseudo Nambu-Goldstone boson. The model is described in detail in arXiv:2412.14243.
}

\section{A parallel universe}

The flavour structure of the Standard Model (SM) looks enigmatic. Both, quarks and charged leptons are arranged into three families with hierarchical increasing masses. However, while quark mixing elements, described by the CKM matrix, are suppressed, analogous mixing elements in the PMNS matrix in the lepton sector are $O(1)$. These structures seem to ask for an underlying unknown dynamics.
Before presenting a possible UV completion of the SM, candidate for this dynamics,\cite{Lizana:2024jby} I would like to start discussing a toy model that may be more familiar for the reader and captures the same underlying idea.

Imagine a parallel universe with only one generation of fermions and without the Higgs boson. This scenario, together with others, is carefully analysed in detail in Ref.\cite{Quigg:2009xr} but here we summarise the main conclusions. A first obvious difference with the real universe would be that the electroweak (EW) gauge symmetry wouldn't be broken at $\sim 100\,$GeV. Would weak bosons remain massless? Not quite, as QCD causes quark pairs to confine at the $\Lambda_{\rm QCD}$ scale $\langle \bar u_L u_R+\bar d_L d_R \rangle \neq 0$ breaking the EW symmetry. The three QCD pions would be eaten by the $W^{\pm}$ and $Z$ bosons, getting a mass of the order of the pion decay constant, $f_{\pi}\sim 100\,$MeV. Below this scale, this toy universe would only be made of electrons and positrons interacting with QED, and neutrinos. Would electrons get a mass? For this, some interaction between the QCD condensate and the leptons is necessary. For instance, if some UV dynamics generates four-fermion operators of the form
\begin{equation}
\mathcal{L} \supset \frac{1}{\Lambda_{\rm UV}^2}(\bar \ell_L e_R)(\bar d_R \, q_L ),
\end{equation}
electrons would get a tiny mass $m_e\sim f_{\pi}^2 \Lambda_{\rm QCD}/\Lambda_{\rm UV}^2$. This UV dynamics could be a Higgs with a positive mass squared or a leptoquark from a Pati-Salam unification. Now, imagine some kind of beings capable of doing physics in this universe, with a technology limited to only energies below $\sim 100\,$MeV. They would describe their universe as QED with electrons. However, they might wonder why the mass of their electrons is so small compared to Planck scale, in a kind of a toy version of the flavour puzzle. Of course, having small masses for fermions is technically natural, but we know there is a deeper reason in this case: although QED is vector-like and it allows for fermion masses, its UV completion, this toy SM, is chiral and forbids them.
It is interesting to also note that, at scales $\sim 100\,$MeV, this version of the SM would naively seem gauge anomalous: quarks are required for gauge anomaly cancelation due to their hypercharges. However, they confine, pions are eaten, and the lightest quark bound states would be the $\rho$ or the $\eta^{\prime}$ mesons, with masses well above $100\,$MeV (for instance, $m_{\rho} \sim 775\,$MeV).

\section{Flavour hierarchies and extended groups}
\label{sec:FHEG}

Perhaps the flavour puzzle of our real SM is addressed by similar dynamics. The role of chirality of the previous example would be played by flavour here. At energies $\sim 100\,$GeV we have the SM, a flavour-universal gauge theory. This is similar to QED being vector-like in the toy universe above.
The Higgs boson, however, couples to the different families breaking strongly flavour-universality and generating flavour hierarchies. Perhaps, the gauge sector of the next layer of physics is flavour non-universal, and this non-universality explains the hierarchies we observe, similarly to the SM being chiral explains the tiny electron masses of the toy universe. This is precisely the idea behind flavour deconstruction, studied in multiple models in the recent years.\cite{Bordone:2017bld,Capdevila:2024gki,Covone:2024elw} In them, the SM group is split into several factors in the UV, broken to the diagonal group to recover the SM. After this breaking, flavour-universality of the gauge interactions emerges in the IR.
Charging the SM fermions and the Higgs into the different factors, one can allow or forbid the different Yukawa couplings, providing a first step towards a dynamical explanation of flavour. 
However, typically these models, to avoid gauge anomalies, treat complete families equally. This is normally translated into similar hierarchical patterns in the quark and lepton sectors, unless extra structure is added.
We here present a model that takes one step further in the analogy with the toy universe above to overcome this issue. The different patterns of CKM and PMNS matrices could come from an anomalous-looking arrangement of the SM fermions into the different gauge groups, triggered by the charges of the fermions of a confining new sector whose condensation breaks the UV symmetry to the SM.\cite{Fuentes-Martin:2024fpx}

Our starting point is gonna be the left-right unification. The EW gauge group $SU(2)_L\times U(1)_Y$ can be embedded into the global group $SU(2)_L\times SU(2)_R\times U(1)_{B-L}$, where $U(1)_Y={\rm diag}[SU(2)_{R}\times U(1)_{B-L}]$. We UV complete this group to $G_{2^41}=SU(2)_{L1}\times SU(2)_{L2}\times SU(2)_{R1}\times SU(2)_{R2}\times U(1)_{B-L}$, but only gauge $G_{2^31}=SU(2)_{L1}\times SU(2)_{L2}\times SU(2)_{R2}\times U(1)_{X}$, where $U(1)_X={\rm diag}[SU(2)_{R1}\times U(1)_{B-L}]$, with $X=T_{R1}^3+\frac{1}{2}(B-L)$. How the SM fermions are split into these factors will be discussed below.

To play the role that QCD has in our toy universe, we add an hyper-coloured sector, with a gauge group $SU(N_{\rm HC})$, and four hyper-fermions in the fundamental representation of each of the $SU(2)$ factors of $G_{2^41}$, $\zeta_L^{(L)}\sim {\bf 2}_{L1}$, $\zeta_R^{(L)}\sim {\bf 2}_{L2}$, $\zeta_L^{(R)}\sim {\bf 2}_{R1}$ and $\zeta_R^{(R)}\sim {\bf 2}_{R2}$. We will assume they transform in the fundamental of $SU(N_{\rm HC})$, but other representations could also be considered. When the hyper-sector confines, pairs of hyper-quarks condensate, $\langle \bar \zeta^{(L)}_L \zeta^{(L)}_R+\bar \zeta^{(R)}_L \zeta^{(R)}_R \rangle \sim f^2 \Lambda_{\rm HC} $, breaking the gauge symmetry $G_{2^31}$ to $SU(2)_L\times U(1)_Y$. Here, $\Lambda_{\rm HC}$ is understood as the scale of the resonances of the hyper-sector, and $f$ the decay constant of the breaking. Using large $N$ arguments, one expects $f\approx \Lambda_{\rm HC} \sqrt{N_{\rm HC}}/4\pi$.

In the SM, the electric charge is the sum of $B-L$ and left and right isospin quantum numbers, $Q=\frac{1}{2}(B-L)+T_L^3+T_R^3$. In the context of our model, we could generalise this formula to $Q=\frac{1}{2}(HB+B-L)+T_L^3+T_R^3$, where $HB$ represents the hyper-baryon number of the new confining sector.
The way to do this is to assign a $HB$ number of $1/N_{\rm HC}$ to hyper-quarks, and to promote $U(1)_{B-L}$ in $G_{2^41}$ to  $U(1)_{HB+B-L}$ so the $U(1)_X$ charges become
\begin{equation}
X=T_{R1}^3+\frac{1}{2}(HB+B-L).\label{eq:Xcharges}
\end{equation}
This is not a wild hypothesis but at this level could seem unjustified. We will see a posteriori that this charging has precisely the effect of addressing the different patterns that the CKM and PMNS matrices have.

We now consider the SM fermions, including three right-handed (RH) neutrinos.
Given this setup, there are nine arrangements of the SM fermions under the $SU(2)$ factors that cancel perturbative gauge anomalies. Among them, only two uniquely identify a subset of a single family of fermions, that we can use as definition of third family fields. 
We explore here one of them because it is the one that will allow for third-family Yukawa couplings in a minimal way. It is illustrated in Table~\ref{tab:arrangement} where we indicate which $SU(2)$ factor the several fermions are doublets of. Table 1 of Ref.\cite{Lizana:2024jby} gives in detail the fermionic representations under all group factors. Checking that this configuration is free of perturbative gauge anomalies is straightforward. The vector character of the hyper-sector makes $SU(N_{\rm HC})^3$ anomalies to cancel, similarly to $SU(3)_c^3$ anomalies. Also, $U(1)_{HB+B-L}$ is vector-like, so there are no $U(1)_{HB+B-L}^3$ or gravitational anomalies. Possible mixed anomalies between the $SU(2)$ factors and $U(1)_{HB+B-L}$ cancel because the sum of $U(1)_{HB+B-L}$ charges of the fields of every cell in Table~\ref{tab:arrangement} vanishes.
Finally, the cancelation of the Witten anomaly imposes that there should be an even number of doublets for every $SU(2)$ factor, implying that $N_{\rm HC}$ must be odd.
This makes $G_{2^41}$ anomaly-free, and thereby, any subgroup of it, in particular, our gauge group $G_{2^31}$. 
Note that it has been fundamental the contribution of the hyper-quarks for the cancelation of mixed anomalies while the gauge uniquely identifies one left-handed (LH) doublet of quarks and one RH doublet of leptons. This is at the core of the explanation of the different hierarchies of the CKM and PMNS matrices.\cite{Antusch:2023shi}

\begin{table}[t]
\renewcommand{\arraystretch}{1.5}
\caption[]{Arrangement of the SM fermions under the four $SU(2)$ group factors of $G_{2^41}=SU(2)_{L1}\times SU(2)_{L2}\times SU(2)_{R1}\times SU(2)_{R2}\times U(1)_{HB+B-L}$. Site $i$ represents both $SU(2)_{Li}$ and $SU(2)_{Ri}$. Remember we only gauge the subgroup $G_{2^31}=SU(2)_{L1}\times SU(2)_{L2}\times SU(2)_{R2}\times U(1)_{X}$, where $X$-charges are given in Eq.~\ref{eq:Xcharges}. SM RH fields are grouped into $SU(2)_R$ doublets, $q_R=(u_R,d_R)$ and $\ell_R=(\nu_R,e_R)$, and we have introduced RH neutrinos. Red (blue) fields are in the fundamental of $SU(3)_c$ ($SU(N_{\rm HC})$).}
\label{tab:arrangement}
\vspace{0.4cm}
\begin{center}
\begin{tabular}{|c||c|c|}
\hline
 & Site 1 & ~~~~~~Site 2 ~~~~~~ \\
\hline
\hline
$SU(2)_L$  & ${\color{red}q_L^{1,2}},~\ell_L^{1,2,3}~,{\color{blue} \zeta_L^{(L)}}$ & ${\color{red} q_L^3}~,{\color{blue} \zeta_R^{(L)}}$  \\
\hline
$SU(2)_R$ & ${\color{red}q_R^{1,2,3}},~\ell_R^{1,2}~,{\color{blue} \zeta_L^{(R)}}$ & $\ell_R^3~,{\color{blue} \zeta_R^{(R)}}$ \\
\hline
\end{tabular}
\end{center}
\end{table}

\section{Composites}
\label{sec:Composites}

The fermionic degrees of freedom of the hyper-sector have a larger global symmetry than $G_{2^41}$: $G_{\rm Global}=SU(4)_1\times SU(4)_2 \times U(1)_{HB}$. The groups $SU(4)_1$ and  $SU(4)_2$ act as complex rotations on $(\zeta_L^{(L)},\zeta_L^{(R)})$ and $(\zeta_R^{(L)},\zeta_R^{(R)})$ respectively with $SU(2)_{L1}\times SU(2)_{R1} \subset SU(4)_1$ and $SU(2)_{L2}\times SU(2)_{R2} \subset SU(4)_2$.
We will take the hyper-sector to confine at scales slightly above the TeV. Then, hyper-quark condensation will break $SU(4)_1\times SU(4)_2 \to SU(4)_V$ giving 15 Goldstones, arranged as $({\bf 1},{\bf 1})+2\times ({\bf 2},{\bf 2})+({\bf 3},{\bf 1})+({\bf 1},{\bf 3})$ of $SU(2)_L\times SU(2)_R$.
Since this breaking contains the gauge braking $G_{2^31}\to SU(2)_L\times U(1)_Y$, the two triplets are eaten by new massive gauge bosons, that under $SU(2)_L\times U(1)_Y$ are a triplet ${\cal W} \sim {\bf 3 }_0$, a neutral singlet ${\cal B}_0 \sim {\bf 1}_0$, and a charged singlet ${\cal B}_1 \sim {\bf 1}_1$. In this case, unlike in the toy universe, there are some surviving physical pseudo Nambu-Goldstone bosons (pNGB): one singlet $S$ and two bi-doublets with the same quantum numbers than the Higgs boson, $H_{1,2}$. A candidate for the Higgs emerges naturally in this setup. To really ensure one of these pNGB can be the Higgs, we need to check how its potential and Yukawa couplings are generated. This will show how the flavour hierarchies emerge.

\subsection{Yukawa couplings}

The SM Yukawas should appear from the interaction of the hyper-sector with the SM fermions, which at this level is absent.
We need to extend our model for this. We will stay agnostic about the particular extension, but let us assume this generates four-fermion operators with the generic form
\begin{equation}
\mathcal{L}\supset  \frac{1}{\Lambda_3^2}\,(\bar \psi_{L} \psi_{R}) (\bar \zeta \zeta ),\label{eq:4fer}
\end{equation}
where $\psi$ is some of the SM fermions. This generates Yukawa couplings with the pNGB Higgses $O(f \Lambda_{\rm HC}/\Lambda_3^2)$ when the hyper-colour confines. Note however that the gauge symmetry restricts the possible SM fermions that we can write in these operators. To generate a SM Yukawa coupling, these operators have to include $SU(2)_L$ and $SU(2)_R$ SM fields with Lorentz indices contracted forming a scalar. Gauge symmetry then forces the hyper-quarks to transform under the same $SU(2)$-factors than the SM fermions. Also, to satisfy Lorentz and $SU(N_{\rm HC})$ symmetry, one hyper-quark must be LH and the other RH. These conditions imply that only SM fields in the diagonals of Table~\ref{tab:arrangement} can get a Yukawa with the pNGB Higgses. These are, Yukawa couplings involving $q_L^3$ and $\ell_R^3$. 
These Yukawas couple to one flavour direction of the complementary chirality, defining the other chirality of third-family fields.
We can thus give mass to the third family:
the gauge symmetry identifies third-family LH quarks and RH leptons, and the Yukawa couplings with these identify third-family RH quarks and LH leptons. 

Before discussing the remaining Yukawas, note that, contrary to $SU(2)_{R1}$ which contains most of the RH SM fields, the $SU(2)_{R2}$ factor containing $\ell_R^3$ is in $G_{2^31}$ and is fully gauged. This implies that the $\tau_R$ has a $\nu^3_R$ partner with same Yukawa coupling with $\nu^3_L$ than the $\tau$, something unacceptable phenomenologically.  However, the model provides a way to get rid of it: we include a singlet fermion $N_L$ which can get a TeV mass with this $\nu^3_R$ when the hyper-sector confines if we include four-fermion operators of the form
\begin{equation}
\mathcal{L}\supset  \frac{1}{\Lambda_N^2}\,(\bar N_{L} \ell^3_{R}) (\bar \zeta^{(R)} \zeta^{(R)}).\label{eq:4ferN}
\end{equation}
$N_L$ and $\nu_R^3$ become a heavy neutral lepton (HNL) which will not participate in the mass generation of the active neutrinos.
Other Yukawas would be generated via higher dimensional operators of the form
\begin{equation}
\mathcal{L}\supset  \frac{1}{\Lambda_{12}^5}\,(\bar \psi_{L} \psi_{R}) (\bar \zeta_L \zeta_{R} ) (\bar \zeta_R \zeta_{L} ).\label{eq:6fer}
\end{equation}
These operators have to involve SM fermions in the same column of Table~\ref{tab:arrangement}, which generate the remaining Yukawas in a suppressed way, $O({f^3 \Lambda_{\rm HC}^2}/{\Lambda_{12}^5})$, due to the higher dimensionality of the operators. Furthermore, in the quark sector, the CKM hierarchy comes from the fact that Yukawas $y_{ij}\bar q^{\,i}_L H q_R^{\,j}$ involving $q_L^{1,2}$ are suppressed with respect to those with $q_L^3$. 
Mixing elements $\theta_{i3}\sim y_{i3}/y_{33}$ are then suppressed.
However, in the lepton sector, gauge symmetry makes no distinction between the three doublets $\ell_L^i$.
Neutrino mass eigendirections are defined via the suppressed Yukawa couplings $y^{\nu}_{ij}\bar \ell^{\,i}_L \tilde H \nu_R^{\,j}$, where $\nu_R^{\,j}$ belongs to $\ell_{R}^{1,2}$. Since the gauge group $G_{2^31}$ does not include the full $SU(2)_{R1}$, but $U(1)_X$, these Yukawas are naturally not aligned to the charged lepton Yukawas, $y^{e}_{ij}\bar \ell^{\,i}_L  H e_R^{j}$, that have a hierarchy imprinted from the RH sector.\footnote{Note that $\nu_R^{\,1,2}$ have a 0 $X$-charge, so they could get a large Majorana mass implementing a seesaw mechanism.}
The PMNS matrix is then naturally anarchic despite charge-lepton hierarchies.
Also, CKM-analogous suppressed mixing angles appear among the RH charged leptons.

\subsection{pNGB potential}

Every breaking of the global symmetry $SU(4)_1\times SU(4)_2$ generates contributions to the potential of the pNGB potential. This includes gauging a subgroup of the global symmetry and the extended sector discussed in the previous subsection. Note that masses of hyper-quarks are forbidden by the gauge symmetry.
The gauge contribution is suppressed with respect to others because it implements a collective breaking: it only gives contributions when the gauging of both factors, $SU(4)_1$ and $SU(4)_2$, is taking into account, so its contribution is doubly suppressed by the gauge couplings of the subgroups of both factors.

Regarding the extended sector, we expect generically it will give a mass $O(f)$ to the pNGBs. Given the phenomenological constraints on the model that we will discuss in the next section, $f\gtrsim 2.5\,$TeV, this would be too large to reproduce the SM. We expect some cancelation could happen for the mass of one of the Higgses to bring it down to the EW scale. This, actually, depends on the particular form of the discussed four-fermion operators. Given the many possibilities, we will use a concrete benchmark here that shows this scenario can appear naturally.
We will only specify the top Yukawa and the HNL mass, that constitute the largest contributions. We assume they are generated via the specific four-fermion operators
\begin{equation}
\mathcal{L}\supset \frac{1}{\Lambda_t^2}(\bar q_{L,\alpha} t_R) (\bar \zeta^{(R)}_{L,1} \zeta^{(L)}_{R,\alpha})+
 \frac{1}{\Lambda_N^2}(\bar  N_{L} \ell^3_{R,\alpha}) (\delta^{\alpha\beta}\bar \zeta^{(R)}_{R,\beta} \zeta^{(R)}_{L,1}+ \epsilon^{\alpha\beta} \bar \zeta^{(R)}_{L,2} \zeta^{(R)}_{R,\beta}),\label{eq:BenchMod}
\end{equation}
where $\alpha,\beta$ are fundamental $SU(2)$ indices and $\epsilon^{\alpha\beta}$ the Levi-Civita tensor.
They could be UV completed if we introduce, for instance, a scalar $\mathcal{S}$, fundamental of hyper-colour and anti-fundamental of colour with $X$-charge $1/2N_{\rm HC}-1/6$, and a scalar $\Phi_R$, bifundamental of $SU(2)_{R1}\times SU(2)_{R2}$. The first one generates the top Yukawa \footnote{We here assume that the mass of ${\cal S}$ is above $\Lambda_{\rm HC}$ so we can integrate it out. If this is not the case, assuming its decay width is smaller than its mass, it will live long enough to confine and form coloured bound states $\bar \zeta \mathcal{S}$. Its couplings with quarks and hyper-quarks generate mixing masses between this resonance and the third-family quarks and then, these SM Yukawa couplings are implemented via partial compositeness.\cite{Kaplan:1991dc} We reach similar conclusions in this case regarding flavour hierarchies.} $y_t\sim \Lambda_{\rm HC} f/\sqrt{2}\Lambda_{t}^2$, while the second one the HNL mass, $M_L\sim 2f^2 \Lambda_{\rm HC}/\Lambda_{N}^2$.
The masses of the physical pNGBs and the quartic of the lightest Higgs $H_1$, $V\supset \lambda_H |H_1|^4/2$, are then
\begin{equation}
m^2_{H^{}_1}\sim \frac{M_N^2-12f^2 y_t^2}{4N_{\rm HC}},~m^2_{H^{}_2}=2m_S^2\sim\frac{M_N^2}{4N_{\rm HC}},~\lambda_H\sim  \frac{3  y_t^2 }{8N_{\rm HC}}.
\end{equation}
We see that the breaking responsible of the HNL mass contributes positively to all pNGB masses giving indeed a mass $O(f)$, but the top Yukawa one contributes negatively only to $H_1$. If both contributions to $H_1$ are similar in magnitude, $M_N^2 \approx 12f^2 y_t^2$, the final mass of $H_1$ will be significantly smaller than $f$. The naive tuning to achieve $m_{H_1}^2 \sim -(100\,{\rm GeV})^2$ while $f \sim (2-3)$TeV is at the percent or per mille level. Furthermore, the magnitude of the quartic coupling falls within the correct range to reproduce the SM Higgs potential.

\section{Phenomenology}
\label{sec:Pheno}

The phenomenology of the model is analysed in detail in Ref.\cite{Lizana:2024jby} We here comment the main features assuming the realisation of Eq.~\ref{eq:BenchMod} for the extended sector.
The most relevant parameters are the decay constant of the breaking $f$, and the two mixing angles of the gauge bosons between gauge and mass bases,
$\sin \theta_L = {g_{L}}/{g_{L2}}$ and $\sin \theta_X = {g_{Y}}/{g_{R2}}$,
where $g_{Li}$ and $g_{R2}$ are the gauge couplings of $SU(2)_{Li}$ and $SU(2)_{R2}$. The SM gauge couplings $g_L,~g_Y$ are related to the UV ones via
$g_L^{-2}=g_{L1}^{-2}+g_{L2}^{-2}$ and $g_Y^{-2}=g_{X}^{-2}+g_{R2}^{-2}$.
The TeV vector boson masses are linked to $f$,
$M_{\cal W}=f(g_{L1}^2+g_{L2}^2)^{1/2}$, $M_{{\cal B}_0}=f(g_{X}^2+g_{R2}^2)^{1/2}$, $M_{{\cal B}_1}={f g_{R2}}$,
and their couplings to the SM fermions are related to the mixing angles. In particular,
the smaller $\theta_{L(X)}$ is, the more strongly $\cal W$ (${\cal B}_{0,1}$) couples to $q_L^3$ ($\tau_R$) and the less to the other SM fermions.

The main observables are LHC searches of the new massive vector bosons, EW precision observables (EWPO) affected by the new vector bosons and the second Higgs, and observables sensitive to the breaking of universality in the quark and lepton sectors. These include observables affected by $bs$ flavour changing neutral currents (FCNC) generated by the ${\cal W}$ such as $B_s$ mixing and $B_s\to \mu\mu$, and charged lepton flavour violation (LFV) processes, triggered by the ${\cal B}_0$, such as LFV $\tau$ decays, $\mu \to  eee$ decays and $\mu\to e$ conversion.
These $\mu \to e$ transitions appear due to the chirally suppressed rotations between light RH leptons and $\tau_R$ that we expect in the model.
Their suppressed effect with respect to LFV $\tau$ decays is compensated by the experimental sensitivity of these observables.

\begin{figure*}[t]
\begin{tabular}{cc}
\includegraphics[width=0.46\textwidth]{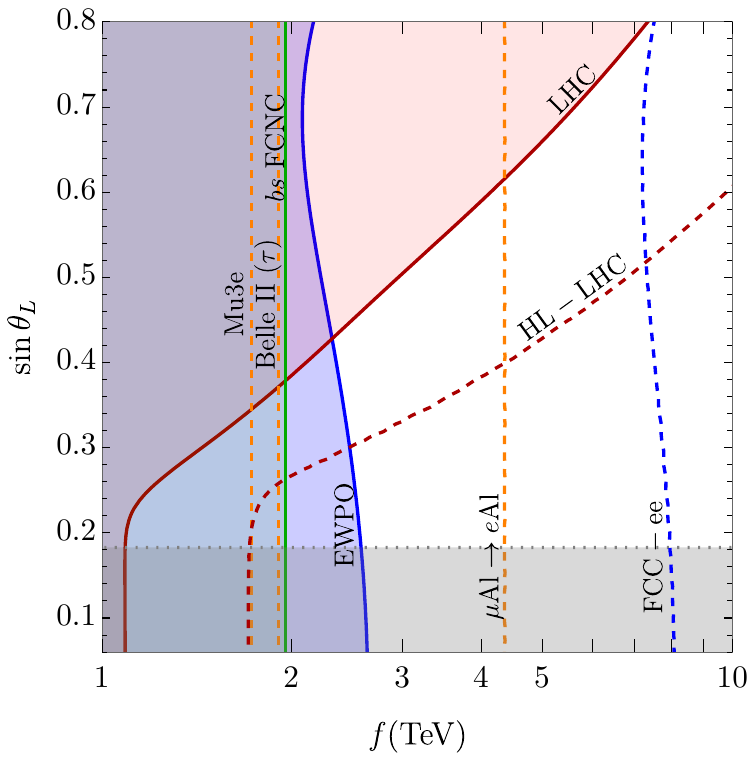} &
\includegraphics[width=0.46\textwidth]{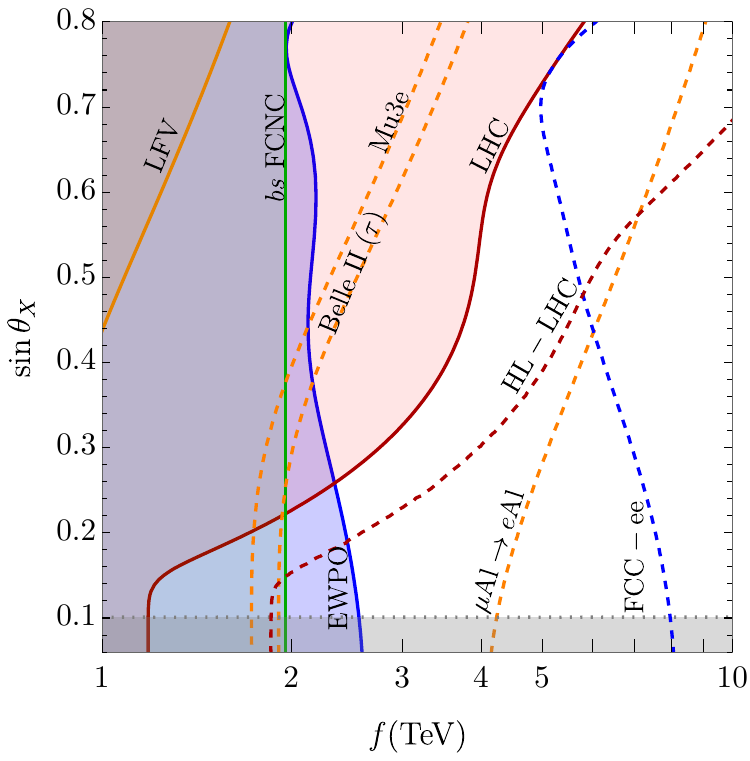} 
\end{tabular}
\caption{Exclusion limits fixing $\sin(\theta_X)=g_Y/2.5$ (left) and $\sin(\theta_L)=g_L/2.5$ (right). Coloured regions are excluded at the $95\%\,$C.L. and dashed lines show projections of future measurements for $95\%\,$C.L. limits. 
For $bs$ FCNC we assume a very mild alignment between the interaction and down bases: the $30\%$ of the CKM rotation comes from the down sector.
The grey areas depict regions where $g^2/(4\pi)>1$ for some of the gauge couplings. We take 3 d.o.f. for EWPO, FCC-ee, LHC and HL-LHC limits, 2 for current LFV limits, and 1 for the others.}
\label{fig:Pheno}
\end{figure*}

We can see the results in Fig. \ref{fig:Pheno}. Current constraints are dominated by LHC searches for larger mixing angles, and EWPO for smaller mixing angles, giving a minimal experimental bound of $f\gtrsim 2.5\,$TeV.
We also show some future projections. These include the High-Luminosity phase of LHC (HL-LHC) for vector boson searches, the Future Circular $e^+e^-$ Collider (FCC-ee) for EWPO and several LFV tests ($\tau$ decays in Belle II, $\mu\to eee$ in Mu3e, and $\mu\to e$ conversion in COMET and Mu2e). It is quite remarkable the improvement in $\mu\to e$ conversion in aluminium from the future experiments COMET and Mu2e, which is capable of testing scales of our model of $f\sim 4\,$TeV and above. 
This shows that this observable will become an important constraint for TeV physics coupled to $\tau$ if chirally suppressed mixings with light leptons are present.

\section*{Acknowledgments}

This work is supported by the grant CSIC-20223AT023 and the Spanish Agencia Estatal de Investigacion through the grant “IFT Centro de Excelencia Severo Ochoa CEX2020-001007-S”.

\end{document}